\documentclass[useAMS,usenatbib]{mn2e}
\usepackage{graphicx,mathptm}

\title[Geometry calibration]
{Maximum-likelihood astrometric geometry calibration of 
interferometric telescopes: application to the Very Small Array}
\author[Klaus Maisinger et al.]
{Klaus Maisinger, M.P.~Hobson, Richard D.E.~Saunders and Keith J.B.~Grainge\\
Astrophysics Group, Cavendish Laboratory, 
Madingley Road, Cambridge, CB3 0HE, UK}
\date{Accepted ---. Received ---; in original form 9 December 2002}

\pagerange{\pageref{firstpage}--\pageref{lastpage}} \pubyear{2002}

\newcommand{\myvec}[1]{{\mathbfit{#1}}}
\newcommand{\mymatrix}[1]{{\mathbfss{#1}}}
\def\pd#1#2{{\frac{\upartial #1}{\upartial #2}}}
\def\nd#1#2{{\frac{\mathrm{d} #1}{\mathrm{d} #2}}} 
\def\spd#1#2#3{{\frac{\upartial ^2 #1}{\upartial #2 \upartial #3}}}
\def\transp{\mathrm{t}}

\begin{document}

\label{firstpage}

\maketitle

\begin{abstract}
  Interferometers require accurate determination of the array
  configuration in order to produce reliable observations.  A method
  is presented for finding the maximum-likelihood estimate of the
  telescope geometry, and of other instrumental parameters,
  astrometrically from the visibility timelines obtained from
   observations of celestial calibrator sources. The method copes
  systematically with complicated and unconventional antenna and array
  geometries, with electronic bandpasses that are different for each
  antenna radiometer, and with low signal-to-noise
  ratios for the calibrators. The technique automatically focusses on 
  the geometry errors that
  are most significant for astronomical observation.
  We apply this method to observations made with the Very Small Array
  and constrain some $450$ telescope parameters, such as the antenna
  positions, effective observing frequencies and correlator amplitudes
  and phase shifts; this requires only $\sim 1$ h of CPU time on a
  typical workstation.
\end{abstract}


\section{Introduction}

Accurate knowledge of observing frequency, projected baselines and
other geometric parameters is essential if interferometer data are to
be turned into useful images. The accuracy required is typically 
a small fraction of the operating wavelength. For example, 
to keep phase errors below 10 degrees for an interferometer operating
at 30~GHz, it is necessary to know the baseline length to an accuracy
of better than 0.3~mm.

The standard way in which interferometer
geometries are determined is along the following lines.
First, direct measurements are made using, for example, tape measures,
theodolites, and measurements of electrical lengths of electronic
systems. Then, observations of celestial calibrators are carried out. For
example, the observation by a particular pair of antennas of a bright
calibrator of known RA and Dec gives the collimation phase of that
pair (the difference in electrical length of the two arms of that
interferometer).  How this quantity varies, for example, with hour angle and
for a series of calibrators at different declinations, is also
investigated. These standard methods are described in e.g.
\citet{thompson94}, and in these ways a series of corrections to the
directly measured geometry is produced, with the aim of ensuring that the
physical baselines (in wavelengths) used in subsequent data analysis are
correct.

This approach is of proven effectiveness for determining the
geometries of interferometer arrays such as the Very Large Array
(VLA). It is not adequate, however, for interferometers such as the
Very Small Array (VSA, see e.g. \citet{watson02}; \citet{scott02}), which is
designed for imaging the CMB on scales $\ga 10$ arcmin at 30~GHz, and thus has
baselines of length $\la 2$~m. At first sight this may seem surprising, as VSA
baselines (in both metres and wavelengths) are so small. From
the point of view of geometry determination from astrometry, however, the VSA
is different from most other interferometer arrays in two respects.
\begin{itemize}
\item
  The collecting area of VSA antennas is very small and so 
  there are few suitable celestial calibrators. Only
  the brightest calibrators are of use and signal-to-noise 
  is a major issue.
\item
  For most arrays like the VLA, the principal swivel axes of each
  antenna (e.g. altitude and azimuth, or polar and declination)
  intersect, at least nominally. If an axis is incorrectly aligned 
  (e.g. the polar axis
  does not quite point to the Pole), then to first order this has no
  effect on the measured phase of a source, and its effect on source
  amplitude will manifest as a simple pointing correction which a
  standard 5-point observation (at expected source position, then offsets
  North, South, East then West) will provide.  
  In the case of the VSA, the table
  elevation axis and the individual horn rotation axes (see 
  Section~\ref{sec:geovsa}) do
  not intersect by $\sim$ 1~m. This design helps reduce important
  systematic errors and also reduces cost, but it makes astrometric
  geometry determination significantly harder. If an axis is misaligned, then
  in general a phase error is also introduced. The identification of such
  errors is exacerbated by the lack of signal-to-noise.
\end{itemize}
There are also further difficulties associated with determining the
geometry of the VSA, which can be present in other interferometers.
\begin{itemize}
\item
As a result of, for example, the compliance needed for 
fitting horn-reflectors to
cryostats, the position of the electrical centre of each horn will be
slightly different, manifesting as a physical baseline correction and
a pointing correction.
\item
A large observing bandwidth makes it hard to achieve a flat band
pass, so the effective radio frequency becomes antenna dependent.
\item
Unconventional geometry makes it difficult to make a straightforward
interpretation of a set of geometry calibration observations.
\end{itemize}

As a result of all these factors, phase (and amplitude) effects arise
from complicated causes that are particularly hard to determine
because of the lack of very high signal-to-noise calibrators. 
We have therefore 
implemented a maximum-likelihood method to determine 
the geometry, and other telescope parameters, 
astrometrically from the visibility timeline data obtained during
a calibration observation of a sufficiently bright unresolved radio source. 
In the geometric model, we include as
variables all the important dimensions and angles that require accurate
determination, including the positions of the antennas, their
effective observing frequencies and phase shifts on different baselines.
In finding the best-fit geometry, the maximum-likelihood method
correctly allows for the signal-to-noise ratio of the observations 
(on the assumption that the instrumental noise is Gaussian); 
the procedure also automatically guards against
attempting to determine unimportant geometry errors that 
do not in fact much affect astronomical observation, 
since these similarly will not much affect calibrator observations.

In this paper, we focus on the method in its application to the VSA, although
we point out the usefulness of the approach generally to radio and
optical interferometry.
In Section~\ref{sec:geocal}, we present the general maximum-likelihood
method for the geometry calibration. The VSA is
introduced in Section~\ref{sec:geovsa}, where we develop a model of the
telescope as a function of the parameters to be calibrated.  Using
this model, the likelihood method is applied to real VSA observations
in Section~\ref{sec:geoapplications}.  Finally, our conclusions are
presented in Section~\ref{sec:geoconclusions}.

\section{Astrometric geometry calibration}
\label{sec:geocal}

Given a multi-dish interferometer consisting of $N_a$ separate
antennas, with antenna positions denoted by $\myvec{d}_i$ ($i = 1,
\ldots, N_a $) relative to some origin, 
one can measure the correlated signals of any two
antennas along $N_b = \frac{1}{2}N_a(N_a-1)$ independent baselines
$\myvec{D}_j$ ($j = 1,\ldots,N_b$), given in some ground-based
reference frame specific to the telescope.  The phase of the signal is
modulated by the geometric path difference that the incident radiation
experiences between the antennas.  If the baseline is represented by a
vector $\myvec{u}$ in a coordinate system
($uv$-coordinates) such that the component~$u$ of the vector
$\myvec{u} = (u, v, w)$ is measured in the east-west direction, 
$v$~in the north-south
direction and $w$ in the direction of the antennas' pointing centre,
the path difference is given by the projection of the direction
cosines $\myvec{x} = (x, y, z)$ of the source on the sky onto the
baseline components, i.e. by $\Delta = \myvec{u} \cdot \myvec{x}$.
Because the field of view is a section of a sphere, the third
direction cosine is $z = \sqrt{1-x^2-y^2}$.  Assuming a small field
size or a sufficiently compact source, $z \approx 1$.

The response of an interferometer to an extended source can be
obtained by integration over the solid angle of the source.  Let us
denote the sky brightness distribution of the source at the observing
frequency~$\nu$ by $I(x, y, \nu)$.  One defines the complex
(not fringe-rotated) {\em visibility} ${\cal V}(u,v, w, \nu) $ as
\citep[compare][]{thompson94}
\begin{equation}
{\cal V} (u, v, w, \nu)
=\int_{-\infty}^{\infty}
\int_{-\infty}^{\infty} A (x,y,\nu) I(x, y,\nu) e^{2\pi i \frac{\nu}{c}
(ux+vy+w)} {dx\ dy},
\label{eqn:wvis}
\end{equation}
where $A(x,y, \nu)$ is the directional antenna (power) response
pattern or {\em primary beam} (normalised to unity at its peak).  To
simplify the notation, we denote the real part
$\Re[\mathcal{V}(\myvec{u})]$ of the
visibility~$\mathcal{V}(\myvec{u})$ by $\Re (\myvec{u})$ and similarly
$\Im (\myvec{u})$ for the imaginary part.

For point sources at positions~$\myvec{x}_k$ from the pointing
centre, whose brightness distributions are given by $I(\myvec{x}) = S_k\ 
\delta (\myvec{x}-\myvec{x}_k)$, the integral~(\ref{eqn:wvis}) for the
complex visibilities~$\mathcal{V}(\myvec{u},\nu)$ reduces
to
\begin{eqnarray}
\Re (\myvec{u}, \nu) & = & \sum_k A(\myvec{x}_k, \nu ) S_k \cos \left(2 \pi
  \frac{\nu}{c} \myvec{u} \cdot \myvec{x}_k \right),\nonumber\\
\Im (\myvec{u}, \nu) & = & \sum_k A(\myvec{x}_k, \nu ) S_k \sin \left(2 \pi
  \frac{\nu}{c} \myvec{u} \cdot \myvec{x}_k \right),
\end{eqnarray}
where the sum is over all relevant sources in the field of view and
$\myvec{x}_k$ and $S_k$ denote, for the $k$th source, respectively the position
from the pointing centre and the flux.  

As the Earth rotates, the $\myvec{u}$-vector corresponding to a
baseline~$\myvec{D}_j$ varies with the time~$t_i$ at which the
visibility sample was measured.  We will denote the $uv$-coordinates
of the $j$th baseline at a time~$t_i$ (or alternatively at hour angle
$H_i$) by $\myvec{u}_{ij} \equiv \myvec{u} (\myvec{D}_j, t_i)$, and
the corresponding visibility by $\mathcal{V}_{ij}$.   The transformation
$\mymatrix{R} (\delta, H_i, L)$ that relates a baseline vector
$\myvec{D}_j$ in the telescope frame to the $uv$-coordinates is a
simple rotation given by
\begin{equation}
\myvec{u}_{ij} = \mymatrix{R} (\delta, H_i, L)\  \myvec{D}_j,
\label{eqn:uvrotate}
\end{equation}
which will generally depend on the geographic latitude~$L$ of the
telescope and the declination~$\delta$ and hour angle~$H_i$ of the
phase-tracking centre.

For a given a model of the telescope, the signal timelines for an
observation of a point source can be predicted and compared with the
observed data. If the source is sufficiently bright, one may then use
a maximum-likelihood approach to determine the antenna geometry and a
large number of other telescope parameters.  For Gaussian instrumental
noise the maximum-likelihood solution for the parameters can be found
by minimising the standard $\chi^2$-misfit statistic between the predicted
visibilities~$\mathcal{V}^\mathrm{P}$ and observed
visibilities~$\mathcal{V}^\mathrm{O}$, namely
\begin{eqnarray}
\chi^2 (\mathcal{V}_{ij}^\mathrm{P}, \mathcal{V}_{ij}^\mathrm{O}) &=&
\sum^{N_t,N_b}_{i,j} w_{ij}\  \left[(\Re^\mathrm{O}
  (\myvec{u}_{ij})-\Re^\mathrm{P}(\myvec{u}_{ij}))^2 \right. \nonumber\\
&&\qquad\,\,\quad
\left. 
+ (\Im^\mathrm{O}(\myvec{u}_{ij})-\Im^\mathrm{P}(\myvec{u}_{ij}))^2 \right]. 
\label{eqn:geomchi2}
\end{eqnarray}
In this expression, the sums run over all $N_t$ time samples 
and $N_b$ baselines.
The weights on the observed visibilities $\mathcal{V}_{ij}^\mathrm{O}$
are given by $w_{ij} = 1 / \sigma_{ij}^2$ assuming that the noises are
independent, which is true for the thermal Johnson noises from each antenna.
For a given observation, $\chi^2$ is a function of
$N_p$~parameters $\{ p_l \}$ ($ l=1,\ldots,N_{p}$), including, for example, the
antenna positions on the table~$\{ \myvec{d}_i \}$, the correlator
amplitudes~$\{C_j\}$, (constant) residual phase shifts~$\{\phi_j\}$
and effective observing frequencies $\{ \nu_i \}$.  By minimising the
$\chi^2$-function with respect to these parameters, it is possible to
find the maximum-likelihood estimate of the telescope geometry.

Finding the global minimum of the $\chi^2$-function is numerically quite
challenging.  The optimisation takes place in an $N_p$-dimensional
space and, as outlined for the VSA below, one typically has
$N_p \approx 450$. Moreover, the $\chi^2 = \mbox{constant}$ hypersurfaces 
are rather uneven.  In particular,
there are many local minima, and a straightforward (e.g. gradient
search) minimisation algorithm may never
find the correct basin of attraction.  The existence of local minima can
be understood intuitively, since a minimum occurs whenever the fringe
pattern is fitted correctly on only one of the baselines.  Even though
the false minimum can easily be recognised from the poor fit on at
least some of the timelines, a straightforward optimisation would stop
there.  Hence a successful fit requires a good initial estimate.  
Simulations show
that for the VSA the initial antenna positions have to be measured
manually to an accuracy of at least a third of a wavelength, i.e.
roughly 3~mm.  This is feasible with a steel tape measure.  It should
also be noted that there is a certain degeneracy in the problem.  For
example, a rescaling of the baselines is equivalent to a scaling of
the observing frequency.

There are a number of possible numerical minimisation algorithms to
choose from.  Some of them employ gradient information, such as
conjugate gradient algorithms \citep{press92}. 
Derivatives of the $\chi^2$-functional with respect to the telescope
parameters can be found in Appendix~\ref{sec:appendix:geo}.  Simulated
annealing methods \citep{kirkpatrick83} try
to find their way through local minima.  In this application, however,
the number of parameters is very large and the $\chi^2$-surface does
not have any significant large-scale slope, rendering a minimisation
in a large number of dimensions very difficult even for the annealing
method. Given a good initial estimate, we found a comparatively simple 
algorithm, Powell's
method~\citep{press92}, to be very robust and reliable.

\section{The Very Small Array}
\label{sec:geovsa}

The VSA consists of an array of 14~horn-reflector
antennas mounted at variable positions on a table that can be steered
in elevation.  The antennas track individually about their horn axes.
A metallic enclosure serves as a ground shield. The observing
frequency can be varied between 26 and 36~GHz, with a receiver
bandwidth of 1.5~GHz and an effective system temperature of 30--35~K.
The telescope was designed to be able to observe in two different
configurations using antennas of different sizes.
\begin{enumerate}
\item The {\em compact array} consists of conical horn reflector
  antennas each with an aperture diameter of 14.3~cm. The primary beam has
  a FWHM of about $4.6\degr$ at a frequency of 34~GHz.  The CMB
  power spectrum is measured in 10~independent bins of width
  $\Delta\ell \approx 80$ from $\ell \approx 100- 900$. The VSA 
  observed in the compact configuration between August~2000 and
  August~2001.
\item The {\em extended array} is a scaled version of the compact
array. The aperture diameter is 32.2~cm, sampling multipoles $\ell
\approx 350 - 1900$ with a resolution of $\Delta \ell \approx 200$.
The VSA has been observing in the extended configuration since
September~2001.
\end{enumerate}

\begin{figure}
  \begin{center}
    \includegraphics[width=6cm]{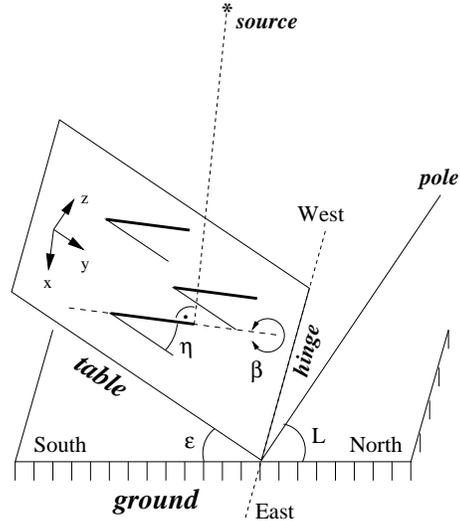}
    \caption{The VSA antennas are mounted on a table with a hinge in
      east-west direction, which tracks a source in elevation
      $\epsilon$. The individual horns are tilted at an angle~$\eta$
      with respect to the table and track in table `azimuth'~$\beta$.}
    \label{fig:vsa_table}
  \end{center}
\end{figure}

The table on which the VSA antennas are mounted can be tilted about a
hinge aligned along the east-west
direction (Fig.~\ref{fig:vsa_table}). The maximum elevation angle
$\epsilon$ of the table is $70\degr$ and the minimum is $0\degr$. 
The antennas have an inclination
relative to the table of $\eta=35\degr$ towards the south, allowing
observations of up to $35\degr$ from the zenith to the north and
south. Each antenna can individually track a source east-west 
up to angles of $\beta=40\degr$ from the meridian.
\begin{figure}
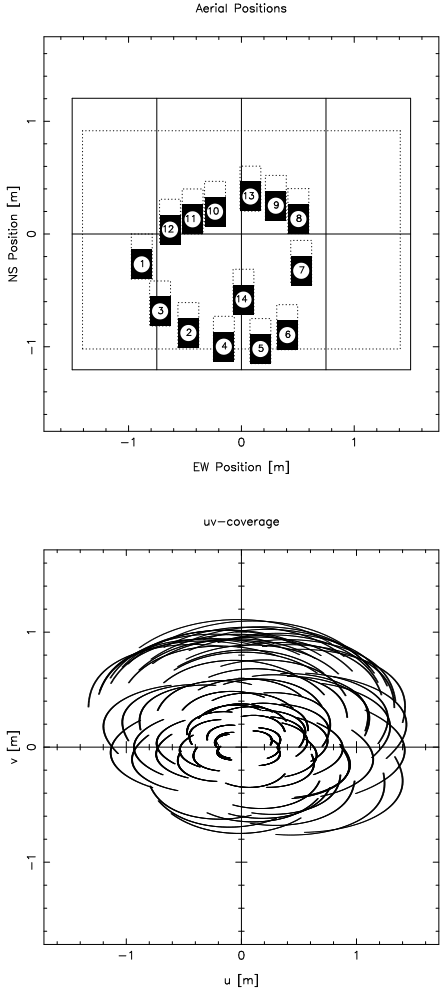

  \begin{center}
    \begin{tabular}{l}
    \includegraphics[angle=-90,width=5.75cm]{antennapos.ps}\\
    \\
    \includegraphics[angle=-90,width=5.75cm]{uv_CRAB.ps}\\
    \end{tabular}
    \caption{Top: the compact arrangement of the 14~VSA antennas on
      the table in September~2000.  The rectangular area indicates the
      shape of the lower part of the VSA table, with the hinge on the
      lower end.  The dashed lines above the horns show the extension
      of the antenna mount \citep{rusholme01}. Bottom: the $uv$-coverage of
      the VSA using the antenna arrangement shown in~(a). The observed
      field is at a declination of $\delta=22\degr$.
      The duration of the observation is 3~hours either side of
      transit.}
    \label{fig:uvcover}
  \end{center}
\end{figure}

For the VSA, the rotation~$\mymatrix{R}$ from~(\ref{eqn:uvrotate}) is a
function of the table elevation~$\epsilon$.  If the elevation has not
been determined accurately, an error~$\Delta \epsilon$ is
introduced. Misalignments of the table hinge can be taken into account
by polar angles $\theta$ and $\phi$ describing the deviation of the
hinge from the east-west direction. Using the three-dimensional
rotation matrices $\mymatrix{R}_x(\phi)$, $\mymatrix{R}_y(\phi)$ and
$\mymatrix{R}_y(\phi)$  for right-handed rotations about the
corresponding coordinate axes, $\mymatrix{R}$ is given by
\begin{eqnarray}
\lefteqn{\mymatrix{R} (\delta, H, L, \Delta \epsilon, \theta, \chi)
  }&&\nonumber\\  
&=&\mymatrix{R}_x(\frac{\pi}{2} - \delta) \mymatrix{R}_z (-H)
\mymatrix{R}_x (L-\frac{\pi}{2})\mymatrix{R}_y (\theta)  \mymatrix{R}_z(\chi) \mymatrix{R}_x
(\epsilon + \Delta \epsilon).
\label{eqn:vsauvrotate}
\end{eqnarray} 

The actual arrangement of the antennas on the table for the compact 
configuration in September~2000 is shown in
Fig.~\ref{fig:uvcover}~(a). The resulting projected baselines during a 6-hour
observation of a field at declination $\delta=22\degr$ are shown 
in panel~(b). Since the telescope table is
stationary with respect to the ground, during a track on a field 
the samples $\myvec{u}$ lie on a series of curves
(or $uv$-tracks).
\begin{figure}
  \begin{center}
    \includegraphics[width=6.5cm]{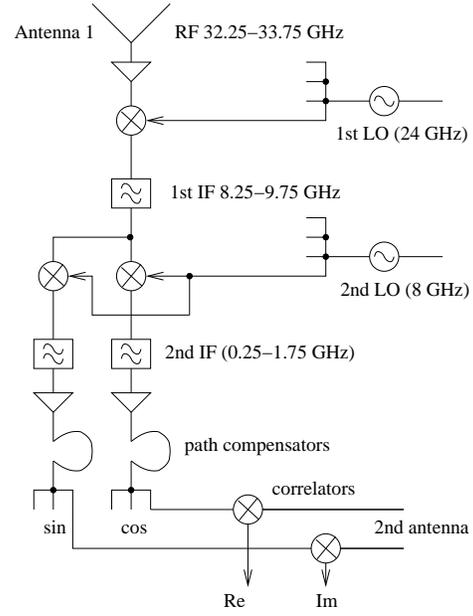}
    \caption{Simplified schematic of the electronics of the receiving
      system of the VSA for compact array observations 
      at a centre frequency of 33~GHz. 
      A signal coming from an antenna at the radio
      frequency $\nu_\mathrm{RF}$ is reduced to the (2nd) IF frequency
      $\nu_\mathrm{IF}$ after mixing with the signals coming from the
      two local oscillators. It is split into $\sin$- and
      $\cos$-channels, and a geometric path delay is inserted into each
      channel separately. The correlated signals from pairs of
      antennas are integrated to produce the real and imaginary parts
      of the visibility.}
    \label{fig:electronics}
  \end{center}
\end{figure}
The electronics of the receiving system of the VSA are illustrated in
Fig.~\ref{fig:electronics}. An antenna receives a signal at the radio
frequency~$\nu_\mathrm{RF}$.  After mixing with the two local
oscillators, which produce a combined frequency $\nu_{\mathrm{LO}} =
\nu_{\mathrm{LO},1}+\nu_{\mathrm{LO},2} $, the frequency of the signal
is reduced to the (2nd) IF frequency $\nu_{\mathrm{IF}} =
\nu_\mathrm{RF}-\nu_{\mathrm{LO}} $.  In order to measure both the
real and imaginary parts of the complex visibility,
the signal is split into 
$\sin$- and $\cos$-channels with a phase difference of 90 degrees,
yielding 28 such channels for $N_a=14$ antennas.  To avoid a loss
of signal coherence due to changing path lengths during a tracking
run, appropriate geometric path delays are inserted into each channel.  
The correlated signals from each pair
of antennas are integrated to produce the
real and imaginary parts of the $N_b$~visibilities.  To minimise the
number of required splits of each $\cos$- or $\sin$-channel, the real
part of the visibility for a given 
baseline can be formed from either the $\cos$-$\cos$ or
the $\sin$-$\sin$ combinations, whereas the imaginary parts may consist
of $\cos$-$\sin$ or $\sin$-$\cos$.  A lookup table is used to determine
which $\cos$- and $\sin$-channels correspond to real and imaginary
parts of a given baseline.

The attenuation and passbands of filters and amplifiers are not
spectrally flat, resulting in slightly different effective IF
frequencies $\nu_{\mathrm{IF}, \cos}$ and $\nu_{\mathrm{IF}, \sin}$
for each antenna.  Additional path delays introduced by the cables and
electronics between antennas and correlators also affect the
visibilities, and the amplitudes of the signal differ due to the
different electronic gains of the correlators.  Therefore the amplitudes
$C^{(r)}_j$ and $C^{(i)}_j$ of the real and imaginary parts and phase
shifts $\phi^{(r)}_j$ and $\phi^{(i)}_j$ vary for each baseline $j$.
However, tests show that these are effectively constant over
timescales of approximately one month on a
given baseline.  

The raw data have to be corrected for inserted path delays
$\rho^{(r)}_{j, l}$ and $\rho^{(i)}_{j, l}$, where $l=1,2$ denotes the
$\cos$ and $\sin$--channels from different antennas corresponding to
the real and imaginary part of the $j$th baseline. The path delays are
introduced at the different IF frequencies $\nu^{(r)}_{\mathrm{IF}, j,
  l}$ and $\nu^{(i)}_{\mathrm{IF}, j, l}$. Note that this notation
with $2\times N_b$ different symbols obscures the fact that there are
just $2\times N_a$ different path delays and IF frequencies, as the
path delays are inserted per antenna and not per baseline.

Gathering together the effects described above, the
predicted visibilities for the $j$th baseline for observations of
a set of point sources at positions $\myvec{x}_k$ with fluxes $S_k$
are thus given by
\begin{eqnarray}
 \Re^\mathrm{P} (\myvec{u}_j) & = & C^{(r)}_j \sum_k A\left(\myvec{x}_k, \sqrt{\nu^{(r)}_{j,1}\nu^{(r)}_{j,2}}\right) S_k \nonumber\\
 &\times& \cos
   \left(2 \pi \left[
   \frac{\sqrt{\nu^{(r)}_{j,1}\nu^{(r)}_{j,2}}}{c} \myvec{u}_j \cdot
   \myvec{x}_k  \right.\right. \nonumber\\
&&\qquad -\left.\left. \left\{\frac{\nu^{(r)}_{\mathrm{IF}, j, 1}}{c} \rho^{(r)}_{j,1} - \frac{\nu^{(r)}_{\mathrm{IF}, j, 2}}{c} \rho^{(r)}_{j,2}\right\} + \phi^{(r)}_j\right] \right),\nonumber\\
 \Im^\mathrm{P} (\myvec{u}_j) & = & C^{(i)}_j \sum_k
 A\left(\myvec{x}_k, \sqrt{\nu^{(i)}_{j,1}\nu^{(i)}_{j,2}} \right) S_k
 \nonumber\\ 
& \times&
\sin
  \left(2 \pi \left[
   \frac{\sqrt{\nu^{(i)}_{j,1}\nu^{(i)}_{j,2}}}{c} \myvec{u}_j \cdot \myvec{x}_k \right.\right.\nonumber\\
 &&  \qquad -\left.\left.\left\{\frac{\nu^{(i)}_{\mathrm{IF}, j, 1}}{c} \rho^{(i)}_{j,1} - \frac{\nu^{(i)}_{\mathrm{IF}, j, 2}}{c} \rho^{(i)}_{j,2}\right\} + \phi^{(i)}_j\right] \right),
\label{eqn:vsavis}
\end{eqnarray}
where $\nu = \nu_\mathrm{IF} +\nu_{LO,1} + \nu_{LO,2}$ is the radio
frequency corresponding to a given IF frequency $\nu_\mathrm{IF}$.
The superscripts~${}^\mathrm{P}$ indicate that the visibilities have been
theoretically predicted. Given the calculable dependence of
$\myvec{u}_j$ for each baseline on the time $t$, one can thus
calculate the quantities $\mathcal{V}_{ij}^\mathrm{P}$ appearing in
the expression for $\chi^2$ given in (\ref{eqn:geomchi2}).

\section{Analysis of VSA calibration observations}
\label{sec:geoapplications}

In order to constrain the geometry, an observation with a good
signal-to-noise ratio at a sufficiently high elevation is required.
There are three possible calibration sources for the VSA compact array
that fulfil these requirements: Cas-A, Cyg-A and Tau-A. These are
essentially unresolved point sources for the VSA compact array.
In the plots presented below, we use a 4-h calibration observation of
Cas-A as an
illustrative example.
The optimisation with respect to antenna
positions, correlator amplitudes and phases and the IF frequencies was
carried out using Powell's algorithm down to a relative convergence
tolerance of~$10^{-3}$ in the value of~$\chi^2$. The entire calculation
required around 1 h of CPU time on a Sparc Ultra workstation.

\subsection{Timelines}

\begin{figure*}
  \begin{center}
    \leavevmode
    \includegraphics[angle=-90,width=\textwidth]{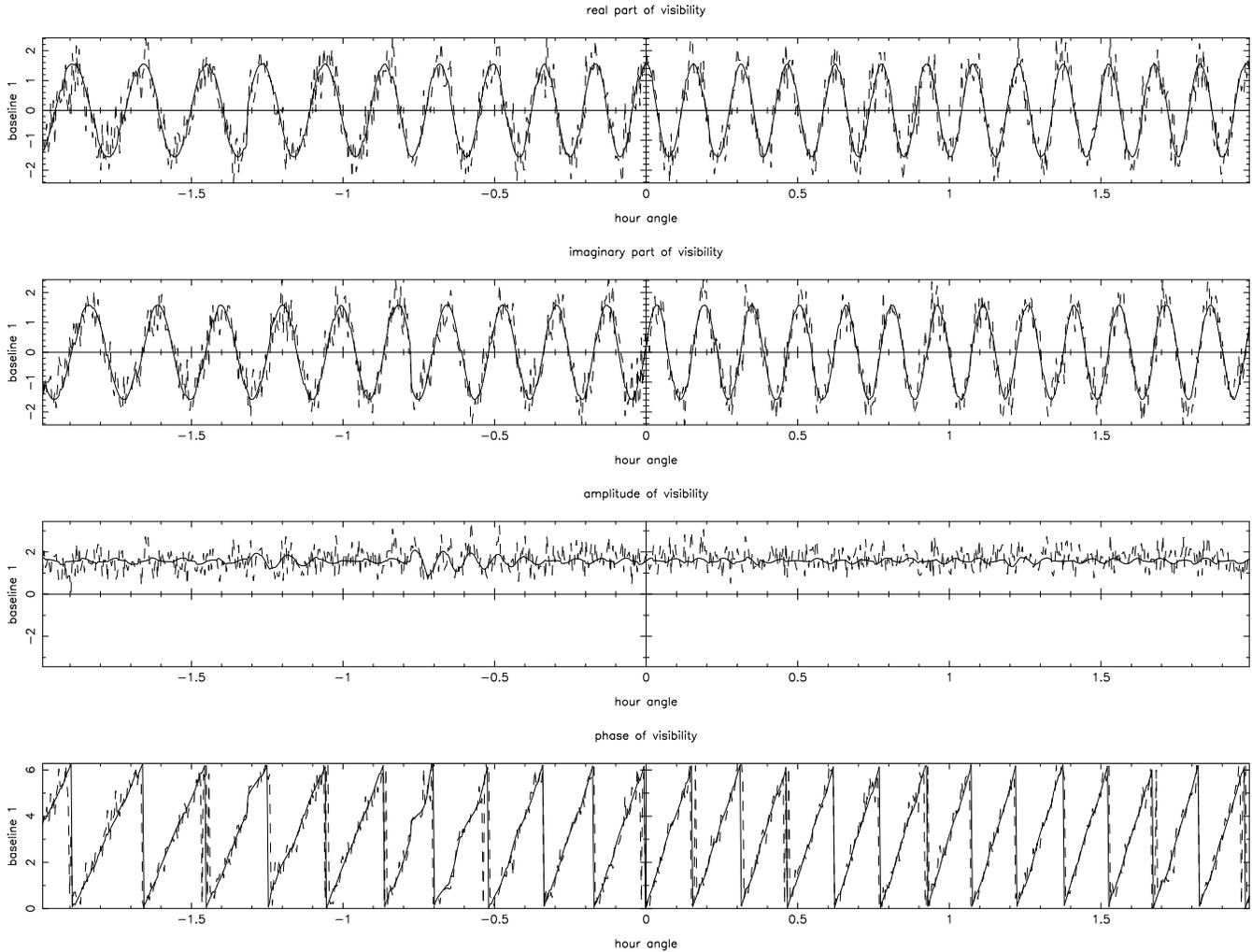}
    \caption{Comparison of observed and predicted fringes on VSA
      baseline antenna 1/antenna 2, 
using an observation of Cas-A.  The panels show
      the real and imaginary parts, the amplitude and the phase of the
      visibility respectively, as a function of hour angle (in hours).
      The theoretical fringe predicted from the optimised telescope
      parameters is plotted with a solid line and the observed fringe
      with a dashed line.  The units on the $y$-axis are arbitrary.}
    \label{fig:base1}
  \end{center}
\end{figure*}

As an illustration of the fit obtained, in Fig.~\ref{fig:base1} we plot
the observed fringes for VSA baseline antenna 1/antenna 2.  
The four panels show respectively
the real and imaginary parts of the visibility, the amplitude and the
phase, each as a function of hour
angle.  Each panel shows the timelines for the observed fringes
(dashed) and the fringes predicted assuming the best-fit telescope
geometry (solid), obtained from the $\chi^2$-minimisation.
The scatter of the observed real and imaginary
parts around the predicted lines indicates the noise level of the
observation.  

The quasi-sinusoidal fringe pattern shows an increasing
fringe frequency for the 1/2 baseline. Since the data have not been 
fringe-rotated, the
phase is going rapidly through cycles of~$2\pi$.
On this baseline, small quadrature errors (see \cite{watson02}) and
differences in the amplitudes on the two channels are seen as a slight
variation of the visibility amplitude and as structure in the
visibility phase. We note that the best fit is
in excellent agreement with the observations and 
matches the observed amplitudes and phases very closely.

\subsection{The $\chi^2$-surface}

\begin{figure}
  \begin{center}
    \leavevmode
    \includegraphics[angle=-90,width=7.12cm]{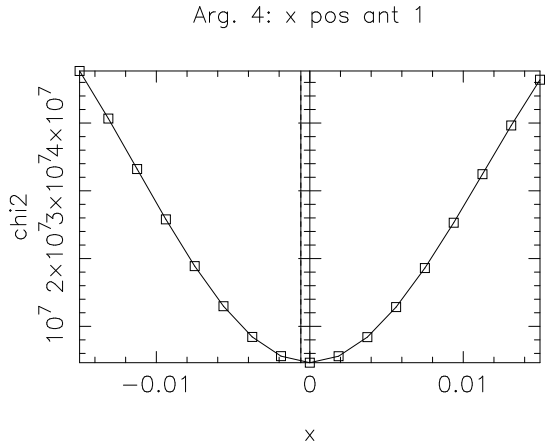}
    \caption{The $\chi^2$-surface as a function of the
      $x$-position of VSA antenna~1. The units on the $x$-axis are
      metres from the optimised position, the $y$-units are arbitrary.
      Low values of~$\chi^2$ correspond to the most likely solution.
      The minimisation has converged to the local minimum in the
      centre. A vertical line indicates the position of the manually
      measured value of the position.}
    \label{fig:geomlikex}
  \end{center}
\end{figure}

\begin{figure}
  \begin{center}
    \leavevmode
    \includegraphics[angle=-90,width=7.12cm]{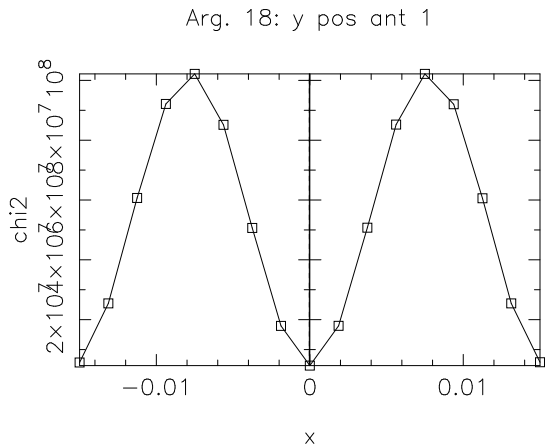}
    \caption{The $\chi^2$-surface as a function of the
      $y$-position of antenna~1. Compare Fig.~\ref{fig:geomlikex}. It
      is interesting to note that adjacent minima are nearly as
      pronounced as the central one.}
    \label{fig:geomlikey}
  \end{center}
\end{figure}

\begin{figure}
  \begin{center}
    \leavevmode
    \includegraphics[angle=-90,width=7.12cm]{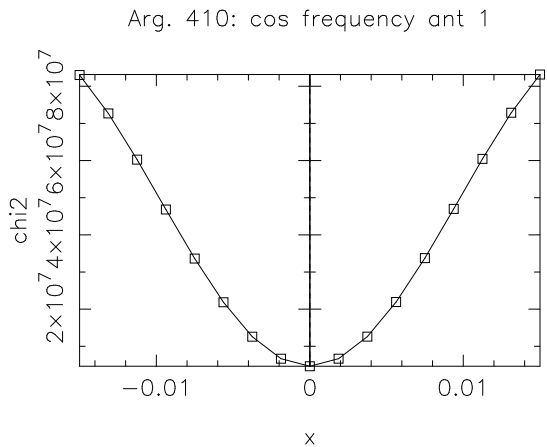}
    \caption{The $\chi^2$-surface as a function of the IF frequency
      on the $\cos$-channel of antenna~1. The $x$-axis is in units of
      12~GHz from the optimal frequency, i.e. 0.01 corresponds to
      120~MHz.}
    \label{fig:geomlikefreq}
  \end{center}
\end{figure}

Confirmation that the optimisation procedure has indeed fully
converged can be seen in Fig.~\ref{fig:geomlikex}, \ref{fig:geomlikey}
and \ref{fig:geomlikefreq}.  These plots show one-dimensional cuts
through the $\chi^2$-surface in parameter space along some of the
parameter coordinate axes.  In each case, the $y$-axis indicates the
value of the $\chi^2$-function, with an arbitrary normalisation, as a
function of the deviation of the parameter value from the best-fit
value. All plots are centred around a local minimum in the
$\chi^2$-surface, indicating that the optimisation procedure has fully
converged, and that the obtained reconstruction is indeed the
maximum-likelihood estimate. There is, however, a strong caveat in
interpreting these one-dimensional $\chi^2$-contours: they are by no
means representative of the accuracy with which the parameters can be
determined, which would require a marginalisation over all other
parameters rather than a simple cut. In fact, the position and shape
of the one-dimensional contour can be changed significantly by simply
varying one or a few of the other parameters. To ensure that a minimum
has been found, all parameters have to converge simultaneously.

Fig.~\ref{fig:geomlikex} shows a cut along the $x$-position of VSA
antenna~1.  Minima in~$\chi^2$ correspond to the most likely solution.
As for all other plots, the minimum has been found. A vertical line
indicates the original input value of the position, as was manually
measured on the table.  Fig.~\ref{fig:geomlikey} shows a cut along the
$y$-position for the same antenna. It is obvious from the figure that,
in this case, the calibration observation used is only just sufficient
to distinguish between the true minimum and the adjacent
minima either side, which are nearly as deep. As mentioned previously,
to avoid being trapped in a false local minimum, one requires a
sufficiently accurate initial estimate.  Nevertheless, it is still possible
to detect false minima, since the resulting positioning mistake of
the order of 1~cm induces significant phase errors on at least one of
the baselines of which the corresponding antenna is a part.

As a further illustration, Fig.~\ref{fig:geomlikefreq} shows a cut
along the IF frequency of the $\cos$-channel of antenna~1.  Once more,
we see that the position obtained by the Powell algorithm lies indeed
at the minimum for this parameter.

\subsection{Imaging with the VSA}

\begin{figure}
  \begin{center}
    \leavevmode
    \includegraphics[width=8cm]{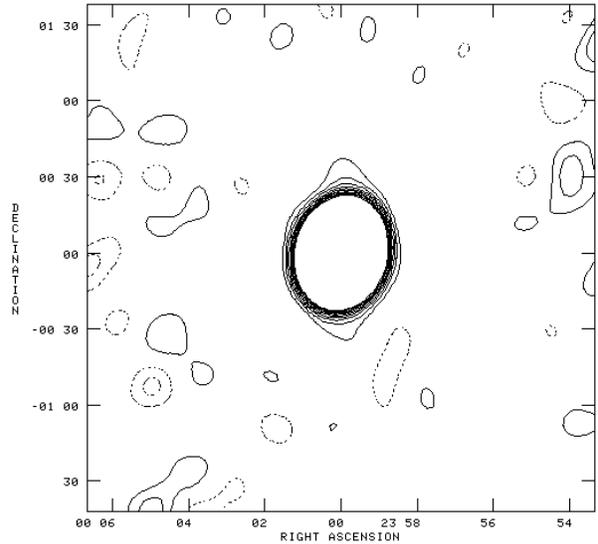}
    \caption{An image of Jupiter observed by the VSA. The geometry and 
      other telescope parameters have been simultaneously calibrated
      by observations of Tau-A and Cas-A. The dynamical
      range of the image is~500:1, indicating that the phase errors
      due to the calibration are indeed negligible.}
    \label{fig:jupiter}
  \end{center}
\end{figure}
The parameters found from the calibration observations can be applied
to the data reduction of subsequent observations.
Fig.~\ref{fig:jupiter} shows a VSA map of Jupiter, which has been
produced using a calibration that combined observations of both Cas-A
and Tau-A. This highlights that the geometry fitting
procedure can easily exploit several data sets simultaneously by
simply adding up the $\chi^2$-functionals for all observations. The
dynamical range of the image is an excellent~500:1, showing that the
geometry calibration on other sources was sufficiently successful to
render phase errors negligible.  Other VSA maps are shown in
\citet{watson02}, and the primordial CMB maps obtained from VSA
observations in the compact configuration are presented by
\citet{taylor02}.

\section{Discussion and conclusions}
\label{sec:geoconclusions}

In this paper, we have presented a method for astrometric geometry
calibration of interferometers by performing a likelihood analysis of
visibility timelines produced during an observation of a reasonably
bright radio point source. The method has been demonstrated for the specific
case of the Very Small Array, for which one must simultaneously
constrain some 450 telescope parameters, such as the antenna
positions, the effective observing frequency of each antenna which can
vary due to different bandpass characteristics, and phase shifts on
different baselines introduced in the electronic system or by
different cable lengths. In particular, analysis of VSA observations
of Tau-A and Cas-A sources show that the relative positions of the
antennas on the telescope table can be determined to a small fraction
of a wavelength. The quality of the fit is sufficiently good to enable
high dynamic-range mapping with the VSA. This astrometric calibration
is very straightforward and cheap, as it requires no equipment other
than the telescope itself.  Multiple calibration observations can
easily be combined to produce a better parameter estimate by summing
the $\chi^2$-functionals for all available data sets.

The calibration method is currently in active use by the VSA project
and has successfully determined the geometry for several different
antenna configuration since the start of its observing programme.  The
calibration of the VSA is described in more detail in
\citet{watson02}, and power spectra derived from the first year of
observations in the compact configuration are presented in
\citet{scott02}.

Although the astrometric calibration method has been applied
specifically to the VSA, it can easily be adapted to other telescopes.
It is worth noting, however, that there is one feature of the VSA
which does not necessarily generalise to other CMB telescopes, even
though it is common to most ground-based arrays.  Each VSA antenna
tracks the source separately in `azimuth' ({\em semi-comounted}
design).  This has the advantage of providing a valuable check on
systematic contaminations, because the path delays on a particular
baseline change during the observation and the sky signal can be
recognised by its characteristic fringe frequency.  On the 
Cosmic Background Imager
\citep{padin02} and the Degree Angular Scale Interferometer \citep{leitch02}, 
the antennas are mounted on a rigid tracking platform
supported by an alt-azimuth mount.  This design keeps the baseline
projections with respect to the source constant and reduces the fringe
rate on all baselines to zero, making the described geometry fitting
impossible.  A similar method can still be applied, however, if the
fringe rate is artificially changed.  For instance, the telescope
could be kept at a fixed pointing position while the source is within
the antenna beam, thus providing a non-zero fringe rate.  
%An
%alternative possibility would be to use a tracking observation with
%constant offset pointing and then to fit for the phase gradients in
%the aperture plane produced by the source.

\section*{Acknowledgments}

We thank Mike Jones, Guy Pooley, Anze Slosar and many 
other members of the VSA
team for their helpful discussions. 
Klaus Maisinger acknowledges support from an EU Marie
Curie Fellowship.

\onecolumn

\appendix
\section{Derivatives used in the VSA geometry calibration}
\label{sec:appendix:geo}

For reference purposes, in this Appendix we quote the 
expressions for the derivatives
of the $\chi^2$-functional~(\ref{eqn:geomchi2}) combined with the
visibilities~(\ref{eqn:vsavis}) used in the maximum-likelihood
algorithm for the the VSA geometry calibration.

The gradient is required by some numerical minimisation algorithms
that use gradient information to accelerate the computation.  Although
not required by the minimisation algorithm, the curvature can be
useful for some iterative minimisation algorithms and   
for quantifying errors or uncertainties on the reconstruction.
Assuming that the maximised estimate of the parameters is indeed in
the correct basin of attraction on the likelihood surface and that
this surface can be reasonably approximated by a Gaussian around the
maximum, the uncertainties in the reconstruction can be obtained from
the inverse of the Hessian matrix of~$\chi^2$.

\subsection{First derivatives of $\chi^2$}
\label{sec:appendix:geo:gradient}

From~(\ref{eqn:geomchi2}), the
$\chi^2$-gradient with respect to the telescope and correlator
parameters~$p_l$ is
\begin{eqnarray}
\pd{\chi^2}{p_l} &= &2 \sum^{N_t,N_b}_{i,j} w_{ij} 
  \left[ \left(\Re^\mathrm{P} (\myvec{u}_{ij})
    - \Re^\mathrm{O} (\myvec{u}_{ij})\right) \nd{\Re^\mathrm{P}
    (\myvec{u}_{ij})}{p_l} + \left(\Im^\mathrm{P} (\myvec{u}_{ij}) - \Im^\mathrm{O}
    (\myvec{u}_{ij})\right) \nd{\Im^\mathrm{P} (\myvec{u}_{ij})}{p_l} \right]
\nonumber
\end{eqnarray}
where again the symbol~$\Re^\mathrm{P} (\myvec{u}_{ij})$ denotes the real part
of the visibility predicted for the $i$th~time sample on the
$j$th~baseline.  The dependence of the visibilities on the parameters
can be explicit, as in the case of the correlator amplitudes, or
implicit in the position of the samples~$\myvec{u}_{ij}$, as for the
antenna positions:
\begin{equation}
 \nd{\Re^\mathrm{P} (\myvec{u}_{ij})}{p_l} = 
\pd{\Re^\mathrm{P}
  (\myvec{u}_{ij})}{p_l} + \pd{\Re^\mathrm{P} 
  (\myvec{u}_{ij})}{\myvec{u}_{ij}} \cdot \pd{\myvec{u}_{ij}}{p_l}.
\end{equation}

We now substitute the expressions~(\ref{eqn:vsavis}) appropriate for
the VSA.  For the sake of notational brevity, we define the arguments
\begin{eqnarray}
\psi^{(r)} &=& \frac{\sqrt{\nu^{(r)}_{1}\nu^{(r)}_{2}}}{c} \myvec{u}_j \cdot
   \myvec{x}_k-\left\{\frac{\nu^{(r)}_{\mathrm{IF}, 1}}{c}
   \rho^{(r)}_{1} - \frac{\nu^{(r)}_{\mathrm{IF}, 2}}{c}
   \rho^{(r)}_{2}\right\} + \phi^{(r)},\nonumber\\
\nu^{(r)} &=&\sqrt{\nu^{(r)}_{1}\nu^{(r)}_{2}},
\label{eqn:abbreviations}
\end{eqnarray}
and similarly for $\psi^{(i)}$ and $\nu^{(i)}$. 
Then the derivatives of the predicted visibilities with respect to the
$uv$-loci, 
\begin{eqnarray}
\pd{\Re^\mathrm{P} (\myvec{u})}{\myvec{u}} & = & 
-2 \pi C^{(r)} \sum_k A(\myvec{x}_k, \nu^{(r)} ) S_k 
\sin \left(2 \pi \psi^{(r)} \right)
\frac{\nu^{(r)}}{c} \myvec{x}_k, \nonumber\\
\pd{\Im^\mathrm{P} (\myvec{u})}{\myvec{u}} & = & 
2 \pi C^{(i)} \sum_k A(\myvec{x}_k, \nu^{(i)}) S_k \cos \left(2 \pi \psi^{(i)} \right) \frac{\nu^{(i)}}{c} \myvec{x}_k,\nonumber\\
\label{eqn:geochi2gradu}
\end{eqnarray}
only depend on table parameters, whereas
\begin{eqnarray}
\pd{\Re^\mathrm{P} (\myvec{u})}{C^{(r)}} & = & 
\sum_k A(\myvec{x}_k, \nu^{(r)} ) S_k 
\cos \left( 2 \pi \psi^{(r)}  \right),\nonumber\\
\pd{\Im^\mathrm{P} (\myvec{u})}{C^{(i)}} & = & 
\sum_k  A(\myvec{x}_k, \nu^{(i)}) S_k \sin \left( 2 \pi \psi^{(i)}  \right),\\
\pd{\Re^\mathrm{P} (\myvec{u})}{\phi^{(r)}} & = & 
-2 \pi C^{(r)} \sum_k  A(\myvec{x}_k,\nu^{(r)} ) S_k 
\sin \left( 2 \pi \psi^{(r)} \right),\nonumber\\
\pd{\Im^\mathrm{P} (\myvec{u})}{\phi^{(i)}} & = & 
2 \pi C^{(i)} \sum_k A(\myvec{x}_k,\nu^{(i)} ) S_k \cos \left( 2 \pi \psi^{(i)} \right),
\end{eqnarray}
only depend on parameters of the electronic system. 

The derivatives with respect to the frequencies are 
\begin{eqnarray}
\pd{\Re^\mathrm{P} (\myvec{u})}{\nu^{(r)}_{\mathrm{IF}, 1}} 
& = & -2 \pi C^{(r)}\sum_k A(\myvec{x}_k,\nu^{(r)} ) S_k \sin \left( 2 \pi \psi^{(r)}
\right)  
\left(\frac{1}{2c} \sqrt{\frac{\nu^{(r)}_{2}}{\nu^{(r)}_{1}}}
  -\frac{\rho^{(r)}_{1}}{c}\right) \myvec{u}_j \cdot
   \myvec{x}_k  ,\nonumber\\
\pd{\Re^\mathrm{P} (\myvec{u})}{\nu^{(r)}_{\mathrm{IF}, 2}} 
& = & -2\pi  C^{(r)}\sum_k A(\myvec{x}_k,\nu^{(r)} ) S_k \sin \left( 2 \pi \psi^{(r)}
\right)  
\left(\frac{1}{2c} \sqrt{\frac{\nu^{(r)}_{1}}{\nu^{(r)}_{2}}}
  +\frac{\rho^{(r)}_{2}}{c} \right) \myvec{u}_j \cdot
   \myvec{x}_k,\nonumber\\
&&
\label{eqn:geochi2gradfreq}
\end{eqnarray}
and analogously for the imaginary parts. These derivatives apply to a
single baseline only.  In practice, a given frequency channel is used
in several baselines.  In order to compute the $\chi^2$-derivatives
one thus has to sum over all baselines containing the respective
frequency channel.

The derivatives of the $uv$-positions with respect to the geometric
parameters are given by
\begin{equation}
\pd{(u_{ij})_k}{p_l} = \left\{  \begin{array}{ll}
\sum_m \pd{R^i_{km}}{p_l} D^j_{m} & \quad \mbox{if $p_l \in \{
\Delta\epsilon, \theta, \phi\}$}\\
\sum_m R^i_{km} \pd{D^j_{m}}{p_l} & \quad \mbox{if $p_l \in \{ \myvec{d}_i \}$}\\
0 \phantom{\pd{D^j_{m}}{p_l}}& \quad\mbox{otherwise}
\end{array} \right.,
\end{equation}
where subscripts denote the vector components, $\myvec{D}$ is a
baseline vector and $\mymatrix{R}(\delta, H_i, L, \Delta \epsilon,
\theta, \chi) $ is the transformation matrix from table to
$uv$-coordinates introduced in~(\ref{eqn:vsauvrotate}):
\[
\myvec{u}_{ij} = \mymatrix{R} (\delta, H, L, \Delta \epsilon, \theta,
\chi) \ \myvec{D}_j(\{\myvec{d}_i\}). 
\]
Derivatives with respect to parameters other than the table parameters
will vanish since they do not affect the geometry.  Furthermore, a
given baseline depends only on the positions of those two horns of
which it is composed, and the resulting matrix is considerably sparse.

\subsection{Curvature of $\chi^2$}
\label{sec:appendix:geo:curvature}

Denoting the real part of the visibility predicted at the $i$th time
sample on baseline~$j$ by $\Re^\mathrm{P}_{ij}$, the $\chi^2$-curvature is
given by
\begin{eqnarray}
\spd{\chi^2}{p_l}{p_n}& =& 2 \sum_{i,j} w_{ij} 
  \left\{ \left( \pd{\Re^\mathrm{P}_{ij} }{\myvec{u}} \cdot
        \pd{\myvec{u}_{ij}}{p_n} \right)
        \left( \pd{\Re^\mathrm{P}_{ij} }{p_l} +\pd{\Re^\mathrm{P}_{ij} }{\myvec{u}}\cdot \pd{\myvec{u}_{ij}}{p_l}
        \right) \right. \nonumber\\  
& & +  \left( \Re^\mathrm{P}_{ij}-\Re^\mathrm{O}_{ij} \right)\ \left[ \spd{\Re^\mathrm{P}_{ij} }{p_l}{p_n} + \left
  ( \pd{\myvec{u}_{ij}}{p_n}\right)^\transp \left[ \nabla_{\myvec{u}}
  \nabla_{\myvec{u}}\Re^\mathrm{P}_{ij} \right]    
    \left ( \pd{\myvec{u}_{ij}}{p_l}\right) +
    \pd{\Re^\mathrm{P}_{ij}} {\myvec{u}}\cdot \spd{\myvec{u}_{ij}}{p_l}{p_n}
  \right] \nonumber\\
&&        + \left(\pd{\Im^\mathrm{P}_{ij} }{\myvec{u}} \cdot
        \pd{\myvec{u}_{ij}}{p_n} \right)
        \left(\pd{\Im^\mathrm{P}_{ij} }{p_l}+\pd{\Im^\mathrm{P}_{ij} }{\myvec{u}} \cdot \pd{\myvec{u}_{ij}}{p_l}
        \right) \nonumber\\
& & + \left.\left( \Im^\mathrm{P}_{ij}-\Im^\mathrm{O}_{ij} \right)\ \left[\spd{\Im^\mathrm{P}_{ij} }{p_l}{p_n}+\left( \pd{\myvec{u}_{ij}}{p_n}\right)^\transp \left[ \nabla_{\myvec{u}}
  \nabla_{\myvec{u}}\Im^\mathrm{P}_{ij} \right]    
    \left ( \pd{\myvec{u}_{ij}}{p_l}\right)+
  \pd{\Im^\mathrm{P}_{ij} } {\myvec{u}} \cdot \spd{\myvec{u}_{ij}}{p_l}{p_n}
  \right] \right\}. \nonumber
  \label{eqn:geochi2curv}
\end{eqnarray}
Again the expressions~(\ref{eqn:geochi2gradu})--(\ref{eqn:geochi2gradfreq})
can be used to substitute for the partial derivatives.

%%%%%%%%%%%%%%%%%%%%%%%%%%%%%%%%%%%%

% \bsp % ``This paper has been produced using the ...''

\label{lastpage}


\begin{thebibliography}{}
\bibitem[\protect\citeauthoryear{Kirkpatrick et al.}{1983}]{kirkpatrick83}
Kirkpatrick S., Gelatt Jr. C.D., Vecchi M.P., 1983, Science,
220, 671
% Optimization by Simulated Annealing
\bibitem[\protect\citeauthoryear{Leitch et al.}{2002}]{leitch02}
Leitch E.M., 2002, ApJ, 568, 28
% Experiment Design and First Season Observations with the Degree
% Angular Scale Interferometer 
\bibitem[\protect\citeauthoryear{Padin et al.}{2002}]{padin02}
Padin S. et al., 2002, PASP, 114, 83
% ``The Cosmic Background Imager''
% in Publications of the Astronomical Society of the Pacific
\bibitem[\protect\citeauthoryear{Press et al.}{1992}]{press92}
Press W.H., Teukolsky S.A., Vetterling W.T., Flannery B.P., 1992, 
Numerical Recipies in Fortran. CUP, Cambridge
% Numerical Recipes in Fortran 77
\bibitem[\protect\citeauthoryear{Rusholme}{2001}]{rusholme01}
Rusholme B., 2001, The Very Small Array. PhD thesis, University of
Cambridge
%
\bibitem[\protect\citeauthoryear{Scott et al.}{2002}]{scott02}
Scott P.F. et al., 2002, submitted to MNRAS, astro-ph/0205380
% First results from the Very Small Array -- III. The CMB power spectrum
\bibitem[\protect\citeauthoryear{Taylor et al.}{2002}]{taylor02}
Taylor A.C. et al., 2002, submitted to MNRAS, astro-ph/0205381
% First results from the Very Small Array -- II. Observations of the CMB
\bibitem[\protect\citeauthoryear{Thompson, Moran \&
    Swenson}{1994}]{thompson94}
Thompson A.R., Moran J.M., Swenson G.W., 1994, Interferometry and
    Synthesis in Radio Astronomy. Krieger Publishing Company, New York 
\bibitem[\protect\citeauthoryear{Watson et al.}{2002}]{watson02}
Watson R.A. et al., 2002, submitted to MNRAS, astro-ph/0205378
% First results from the Very Small Array -- I. Observational methods

\end{thebibliography}
\end{document}